\documentclass{aa}  
\usepackage{txfonts}
\usepackage{graphicx}
\usepackage{gensymb}
\usepackage{multirow, makecell}
\usepackage{pdflscape}
\usepackage{hyperref}
\hypersetup{
    colorlinks=true,
    linkcolor=blue,
    filecolor=blue,      
    urlcolor=blue,
    citecolor=blue,
}
\begin{document}

\title{The \textit{SVOM}/ECLAIRs image trigger with wavelet-based background correction optimised with a one-year simulation of observations}

\titlerunning{The \textit{SVOM}/ECLAIRs image trigger with wavelet-based background correction}

\author{N. Dagoneau\inst{\ref{inst1}} \and S. Schanne\inst{\ref{inst1}}}

\institute{
CEA Paris-Saclay/IRFU, F-91191 Gif-sur-Yvette, France\\\email{nicolas.dagoneau@cea.fr}. \label{inst1}
}

\date{Received XXX; accepted XXX}

\abstract
   {The  Space-based multi-band astronomical Variable Objects Monitor (\textit{SVOM}) mission under development will carry four instruments, and in particular the coded-mask telescope named ECLAIRs, with a large field of view of about 2 sr, operating in the 4–150 keV energy band. The trigger software on board ECLAIRs will search for high-energy transients such as gamma-ray bursts and peculiar behaviour (e.g. strong outbursts) from known X-ray sources, in order to repoint the satellite to perform follow-up observations with the onboard narrow field of view instruments.}
   {The image trigger, one of the two algorithms implemented in the software on board ECLAIRs, produces images over periods of exposure ranging from 20 seconds to 20 minutes   during which the Earth can cross the field of view. The Cosmic X-ray Background and contributions from known X-ray sources are expected to dominate the ECLAIRs astrophysical and instrumental background and must be taken into account and corrected prior to coded-mask image deconvolution in order to optimise the sensitivity to faint transients.}
   {To correct these background components, we implemented and studied a traditional fitting method and a new method based on wavelet decomposition of the detector image. In order to study and to assess the performance of these methods, we performed a one-year simulation of the image trigger on board ECLAIRs. From the images produced during this realistic observation scenario of the \textit{SVOM} mission, we also defined a way to analyse the sky images to search for new sources.}
   {We present the algorithms behind the image trigger on board \textit{SVOM}/ECLAIRs. We show that the wavelet method we implemented provides similar results in terms of cleaning performance compared to the traditional fitting method, and has the benefit of not requiring any assumption on the shape of the background on the detector. 
   We also calibrate the detection threshold to be adaptive and based on the quality of the reconstructed sky image.
   } 
   {}
   

\maketitle
\section{Introduction}

Transient astrophysical events include various high-energy phenomena such as gamma-ray bursts (GRBs) and flares from X-ray binaries or highly magnetized neutron stars. These short-duration phenomena in the X-ray and gamma-ray energy domains deserve  fast detection to permit multi-wavelength follow-up observations, which  will allow us to  study of the underlying physics and to determine the distance of the emitting source.

This strategy will be followed by \textit{SVOM} \citep{wei_deep_2016}, a French-Chinese mission currently under development and planned to be operational after 2022. The fast detection of transient sources needs to be carried out directly on board the satellite since it cannot rely on ground-based data processing due to the lack of real-time large-volume data transmission from low-Earth orbit to the ground.

The goal of the ECLAIRs coded-mask aperture telescope on board \textit{SVOM} \citep{takahashi_x-gamma-ray_2014} is to observe a large portion of the hard X-ray sky and automatically detect and localise GRBs and other kinds of transient sources, thanks to its trigger software \citep{schanne_svom_2019}. The software is implemented in the Scientific Trigger and Control Unit (UGTS, French for \textit{Unité de Gestion et de Traitement Scientifique}) \citep{schanne_scientific_2013,le_provost_scientific_2013}.

The ECLAIRs instrument is made up of the UGTS; the CdTe detector plane (80 $\times$ 80 pixels of active surface 4 $\times$ 4 mm$^2$, 1 mm thick, separated by 0.5 mm), which  detects photons from 4 to 150 keV; and the self-supporting tantalum mask (54 $\times$ 54 cm$^2$ in size and 0.6 mm thick), which provides imaging capabilities up to 120 keV and is  located at a distance of 46 cm above the detection plane.

The UGTS is composed of ten electronic boards that perform different functions (power management, data input-output management, and data processing). Two redundant boards are dedicated to data processing. Each of these includes an FPGA to preprocess the data for the trigger software and a Leon3 dual-core CPU running the complete ECLAIRs flight software, including the trigger.

The trigger software is divided into two different algorithms to reflect the diversity of transient sources in terms of duration. The first is a count-rate trigger, well suited to detecting short GRBs. It monitors the counts recorded by the detector plane on timescales between 10 ms and 20.48 s, and performs excess detection over background and imaging of those excesses. The second algorithm is an image trigger, well suited to detecting long GRBs. It analyses images built on timescales from 20.48 s to approximately 20 min. This image trigger is also well adapted to the detection of ultra-long GRBs \citep{dagoneau_ultra-long_2020}. The present paper focuses on the general description of the image trigger, including its algorithm and the processing, which are implemented to optimise the detection of transient sources in the images that are produced. 

\section{Image trigger}
\label{sec:image_trigger}

\begin{figure*}
\centering
   \includegraphics[width=17cm]{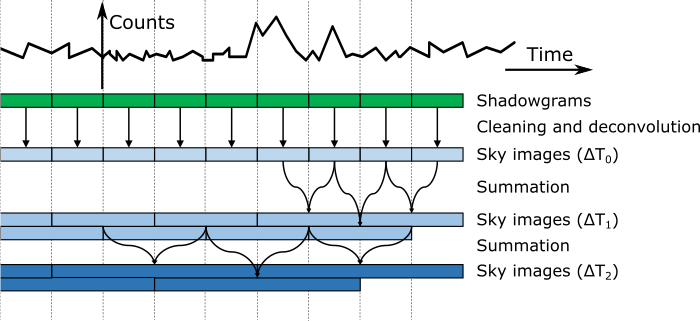}
     \caption{Illustration of the algorithm behind the image trigger for one energy strip. The vertical dotted lines delimit portions of 20.48 s. The green blocks correspond to shadowgrams that are corrected and deconvolved. The blue blocks correspond to the images of the sky. For visibility, only scales $n=0$, $n=1$, and $n=2$ are shown (scale $n=6$ is the result of the sum of 64 images of 20.48 s).}
     \label{fig:trigger_image}
\end{figure*}

The image trigger is the name given to one of the two trigger algorithms implemented in the ECLAIRs onboard detection software. An illustration of this algorithm is given in Fig. \ref{fig:trigger_image}. This algorithm   runs in cycles of 20.48 s in four configurable energy strips to reflect the spectral diversity of the transient sources; in this paper the four strips are 4--20 keV, 4--50 keV, 4--120 keV, and 20--120 keV. The 80 $\times$ 80 pixel images recorded by the detector plane within 20.48 s are called shadowgrams. The sky image is reconstructed from the shadowgram using the mask deconvolution method (as is currently used by the IBIS telescope on board the ESA INTEGRAL observatory; a description of the method can be found in \citealt{goldwurm_integral/ibis_2003}). This method permits the  reconstruction of  point-like sources in the energy band between 4 and 120 keV where the mask (Ta of 0.6 mm thickness) is opaque enough for such sources to produce shadowgrams with good contrast.

The sky image has a size of 200 $\times$ 200 pixels with a pixel angular size ranging from 33 arcmin in the centre  of the field of view to 17.5 arcmin at the edge.
The deconvolution uses the detected number of counts per pixel $D_{\mathrm{cnt}}$ and, assuming a Poissonian distribution per detector pixel ($D_{\mathrm{var}}$ = $D_{\mathrm{cnt}}$), it produces reconstructed sky images in number of counts ($S_{\mathrm{cnt}}$) and variance ($S_{\mathrm{var}}$). 
A signal-to-noise (S/N) sky image is also built: $S_{\mathrm{S/N}} = S_{\mathrm{cnt}} / \sqrt{S_{\mathrm{var}}}$.

Before the deconvolution, the shadowgrams need to be cleaned from the Cosmic X-ray Background (CXB) and the contributions from bright known X-ray sources. The cleaning step is  described in Sect. \ref{sec:cleaning}. The background correction is performed on the detected counts image. In theory, this correction leads to an increase in variance in the shadowgram. However, this increase is negligible (less than $1\%$ in the corner of the shadowgram) compared to the initial number of counts in 20 s (and thus to the initial variance when modelling the distribution of the counts by a Poisson distribution). For this reason, the uncorrected count images are deconvolved to produce the variance of the sky.

The deconvolution leads to sky images that are subsequently added (separate summation of the counts and the variances) into a history of the most recent sky images. The resulting exposure time of the sky images produced by summation is $\Delta T_n=20.48\times 2^{n}$~s with $n=[0,6]$. Thus, the available scales range from $\Delta T_0$ = 20.48 s to $\Delta T_6$ = 1310.72 s (approximately 20 min, see Fig. \ref{fig:trigger_image} for an illustration). Each of the sky images is searched for a transient source. This step is described in Sect. \ref{sec:analyse}.

Unlike the image triggers developed for previous missions like Swift/BAT, the ECLAIRs image trigger must be able to operate when the Earth passes through its field of view (2 sr), which occurs approximately every orbit (90 minutes). The Earth passages are the result of the strategy set up by SVOM to point the ECLAIRs field of view  towards a roughly anti-solar direction in order to trigger on sources that  are immediately observable by ground-based observatories located on the Earth's night hemisphere. The management of the Earth in the field of view is  explained throughout the paper at each step of the image trigger. 

The ECLAIRs telescope will also transmit all the counts recorded by the camera to the ground by X-band, where the detection software can be executed offline (typically with 6--12 hr of delay) to adjust the parameters, detect low significance sources, and permit the update of the onboard software configuration by telecommand.  

\section{Methodology}
\label{sec:method}

\begin{figure}
\resizebox{\hsize}{!}{\includegraphics{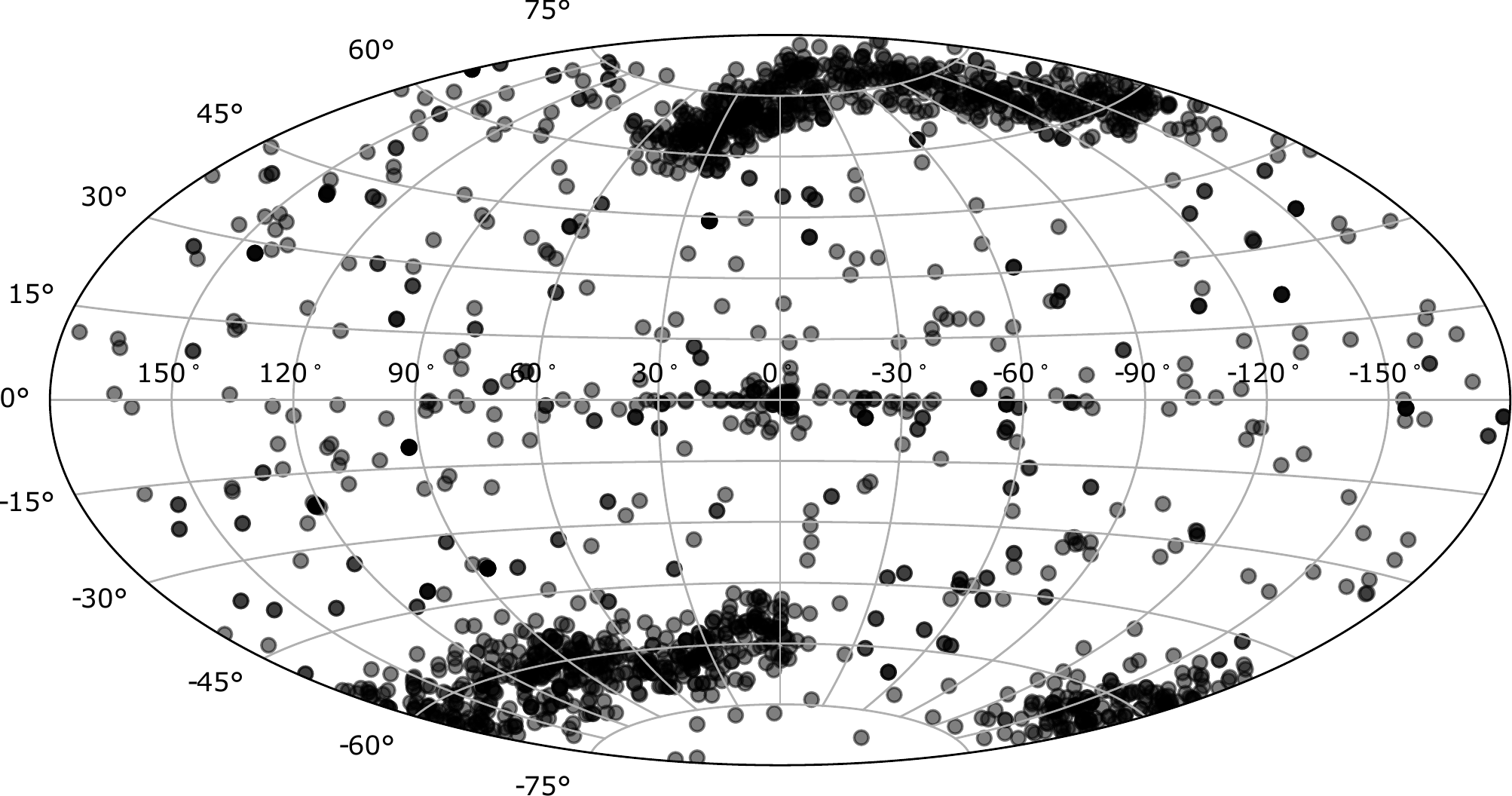}}
\caption{Positions of the pointing directions on the sky in Galactic coordinates in a one-year simulation by CNES (nominal phase, \citealt{jaubert_realistic_2017}).}
\label{fig:pointings}
\end{figure}

In order to develop and study the performance of the image trigger, a simulation of a one-year \textit{SVOM} observation sequence carried out by the CNES  is used \citep{jaubert_realistic_2017}. This simulation gives the satellite's position on its orbit and its attitude for every minute of this year. Figure \ref{fig:pointings} shows the positions of the pointing directions on the sky in Galactic coordinates in the one-year mission simulation. There are 1919 stable pointings.

For each pointing the shadowgrams are drawn by ray-tracing CXB and known sources photons in the field of view. A spatially and spectrally flat noise component is also added to the shadowgram at the level of 0.003 counts/s/cm$^2$/keV (taken from the spectrum given in \citealt{mandrou_wide-field_2008}, Fig. 2, integrated over the ECLAIRs energy band), which takes into account the mean contribution of particles, predominantly electrons, along the orbit outside of the South Atlantic Anomaly region. The catalogue of known sources is presented in \citealt{dagoneau_onboard_2021}. The image trigger processes the shadowgrams. In the following we consider a perfect efficiency of the detector between 4 and 150 keV, and we focus on the imaging process in the ECLAIRs full band of 4--120 keV.

In this paper we present the optimisations of the image trigger in terms of imaging performance, background modelling and subtraction. The simulation of GRBs to estimate the GRB detection rate by ECLAIRs were presented previously  \citep{wei_deep_2016,antier-farfar_detection_2016}, based on a prototype trigger implementation \citep{schanne_scientific_2013}.

In order to evaluate the quality of the S/N sky images produced in which the image trigger searches for new sources, we analysed the standard deviation of the S/N as well as the maximum of the S/N in each image at each timescale. In a cleaned image (i.e. in a sky image built from the deconvolution of a clean shadowgram, and possibly summed with other sky images) the S/N should follow a standard normal distribution $\cal{N}$(0,1).
\section{Before deconvolution: Cleaning}
\label{sec:cleaning}

The deconvolution may lead to some artefacts that mimic point sources when the background noise is not uniform in the shadowgram. Because of the geometry of the ECLAIRs telescope (and in particular the distance between the mask and the detector of 46 cm), an isotropic source such as the CXB manifests itself by a non-uniform distribution of the counts on the detector. The deconvolution of such a shape generates artefacts in the images of the sky that increase the width of the distribution of S/N and reduce the detection efficiency of faint sources. Before the deconvolution, the shadowgrams ($\Delta T = 20.48$~s) must therefore be corrected for the CXB, but also for the contribution of the known bright sources (the effects of not correcting bright sources are presented in \citealt{dagoneau_onboard_2021}). 

\subsection{CXB correction}

To correct the CXB, two methods are implemented in the onboard software (with possible switching by telecommand). The method that will be used by default is the traditional way already used in the past (as an example for the \textit{Granat}/SIGMA data processing, \citealt{bouchet_sigmagranat_2001}), which consists in subtracting a 2D polynomial shape after a fit in the shadowgram. This 2D polynomial model is justified by the geometry of the instrument. For a large field-of-view instrument, an isotropic source in the sky such as the CXB leads to a 2D distribution of the counts in the shadowgram with more counts in the centre than in the corners. This non-flat distribution flattens out when the field of view decreases (e.g. when the distance between the mask and the detector increases). The model is defined by Eq. \ref{eq:2d_shape} where $i$ and $j$ refer to the pixel coordinates in the shadowgram:

\begin{align}
\label{eq:2d_shape}
M(i,j) = c_0 + c_1\cdot i + c_2 \cdot j + c_3 \cdot i^2 + c_4 \cdot j^2 + c_5 \cdot i \cdot j.
\end{align}

In the absence of the Earth in the field of view, this 2D shape appears as a curved shape with a maximum at the centre of the shadowgram.
However, in the case of \textit{SVOM}/ECLAIRs, this method should also operate with the presence of the Earth in the field of view. The 2D shape applied allows for this correction, and permits the modelling of  a deficit of counts in the direction of the Earth (see Fig. \ref{fig:ex_image_fit_wt}) without explicitly injecting into the model the coordinates of the Earth in the field of view.

We have implemented another method using wavelets and the \textit{à trous} algorithm (French for `with holes')  \citep{holschneider_real-time_1990, starck_astronomical_2002}. This is the first time that this algorithm will be used to correct the background in detector images from a coded-mask aperture telescope prior to deconvolution (but   already applied after the deconvolution to remove systematics is sky images,  \citealt{krivonos_integral/ibis_2010}). It results from the fact that the background shape produced by the large size of the field of view modulated by the Earth's presence appears as a large-scale structure on the shadowgram, while the point-source contributions are imprinted in small-scale structures on the shadowgram because of the small size of the coded-mask elements. 

Thus, the shadowgram can be decomposed into different scales and a background-corrected version of the shadowgram is reconstructed using only the smallest scales. The shadowgram $D_s$ at a scale $s \geq 1$ and pixel position $(i,j)$ is computed according to Eq. \eqref{eq:shds_scale_s} which is a convolution of the shadowgram at the previous scale $D_{s-1}$ and a filter $H$; $D_0$ corresponds to the raw shadowgram:

\begin{align}
\label{eq:shds_scale_s}
D_s(i,j) = \sum_{m=-l}^{l} \sum_{n=-l}^{l} H(m,n) D_{s-1}(i+2^{s-1}m, j+2^{s-1}n)
\end{align}

\begin{figure}
\resizebox{\hsize}{!}{\includegraphics{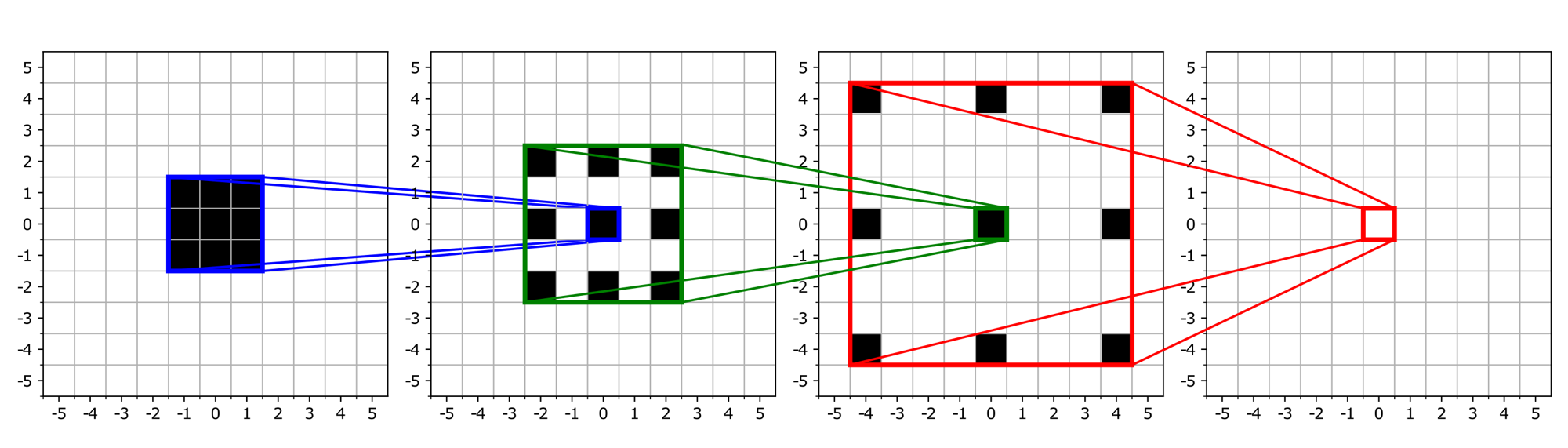}}
\caption{Illustration of the calculation of the approximations by the \textit{à trous} algorithm applied to the central pixel $(i, j) = (0, 0)$ for a filter of size $3 \times 3$. At each scale $D_0$, $D_1$, and $D_2$ (from left to right),    the pixels involved in the convolution with the filter to compute $D_1(0, 0)$, $D_2(0, 0)$, and $D_3(0, 0)$ are shown in black.}
\label{fig:algo_wavelet}
\end{figure}

The algorithm is called \textit{à trous} because, while it calculates the different scales of the detector image, the distance between two pixels considered for the convolution by the filter increases by a factor of 2. To compute the pixels of $D_1$ the pixels taken into account are neighbours in $D_0$, whereas for the calculation at scale s they are separated by a distance $2s-1$, as shown in Fig. \ref{fig:algo_wavelet}. This allows  larger and larger structures to be caught in the shadowgram as the scale increases. To keep the detector image size identical at all scales ($80 \times 80$), symmetric boundary conditions are used: the image of the pixel of index $-1$ (which would be outside the matrix) is the pixel of index 0; the image of the pixel of index 80 (also outside the matrix) is the pixel of index 79. Symmetric boundary conditions are appropriate for sources in the partially coded field of view (in contrast to cyclic boundary conditions).

The 2D filter $H$ is built from a 1D filter $h$ such that $H=hh^T$. There are two possible 1D filters that are commonly used in the \textit{à trous} algorithm \citep{starck_undecimated_2007}: $h_3=(1/4, 1/2, 1/4)$ or $h_5=(1/16, 1/4, 3/8, 1/4, 1/16)$. The wavelet coefficients $W_s$ at a scale $s \geq 1$ and pixel position $(i,j)$ are given by Eq. \eqref{eq:w_s_diff}:

\begin{align}
\label{eq:w_s_diff}
W_s(i,j) = D_{s-1}(i,j) - D_{s}(i,j).
\end{align} 

The image of the original shadowgram $D$ can be recovered by summing the different wavelet coefficients $W_s$ and the last scale $D_{s_{\mathrm{max}}}$ at the largest scale considered $s=s_{\mathrm{max}}$:

\begin{align}
D(i,j) = \sum_{s=1}^{s_{\mathrm{max}}} W_s(i,j) + D_{s_{\mathrm{max}}}(i,j).
\label{eq:rebuilt_wavelet}
\end{align}

The removal of the CXB is achieved by the reconstruction of a cleaned shadowgram $D_{\mathrm{cleaned}}$ summing only a limited number of scales $s=s_{\mathrm{th}} < s_{\mathrm{max}}$. Thus, given that the wavelet coefficient is the difference between two consecutive scales, the cleaned shadowgram can be computed by the subtraction of the shadowgram at scale $s_{\mathrm{th}}$ and the raw shadowgram $D_0$:

\begin{align}
D_{\mathrm{cleaned}}(i,j) = D_0 - D_{s_{\mathrm{th}}}.
\label{eq:rebuilt_cleaned}
\end{align}

Thus, the algorithm needs two parameters to be chosen: one filter among the two presented previously ($h_3$ or $h_5$) and the scale $s_{\mathrm{th}}$ to be used in order to build the corrected shadowgram containing only the smallest scales, which contain most of the point-source coded signal while reducing the influence of the background. The preliminary choice of these parameters is made in such a way that the largest scale of the shadowgram $D_{s_{\mathrm{th}}}$ contains most of the non-uniform background. In order to evaluate this and to deduce the most adapted parameters, we simulated shadowgrams of 20 s containing only CXB (plus the additional flat noise). The shadowgrams are cleaned with the wavelet method using various combinations of the two parameters and then deconvolved to produce the sky images that are summed together to reach an exposure time of 20 min. The best parameters are those  that lead to a distribution of the S/N with a standard deviation as close as possible to 1 in the images of the sky in 20 min. As with the fit method, the uncorrected counts shadowgram is used as the variance to produce the variance of the sky, and thus to compute the S/N.

\begin{figure}
\resizebox{\hsize}{!}{\includegraphics{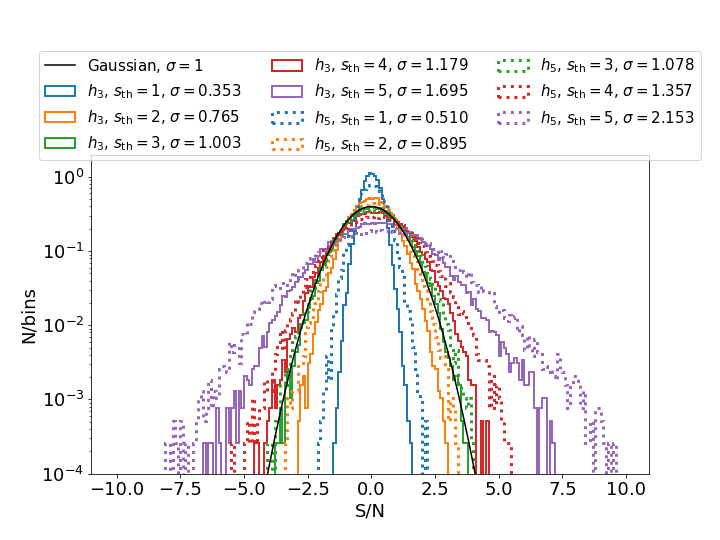}}
\caption{Distribution of the S/N in 20 min sky images for different combinations of the filter size and the maximum scale used for the wavelet cleaning. The black solid curve corresponds to the normal distribution.  The best parameters ($h_3$, $s_{\mathrm{th}}=3$) are shown as the solid green   curve.}
\label{fig:wt_h_sth}
\end{figure}

\begin{figure}
\resizebox{\hsize}{!}{\includegraphics{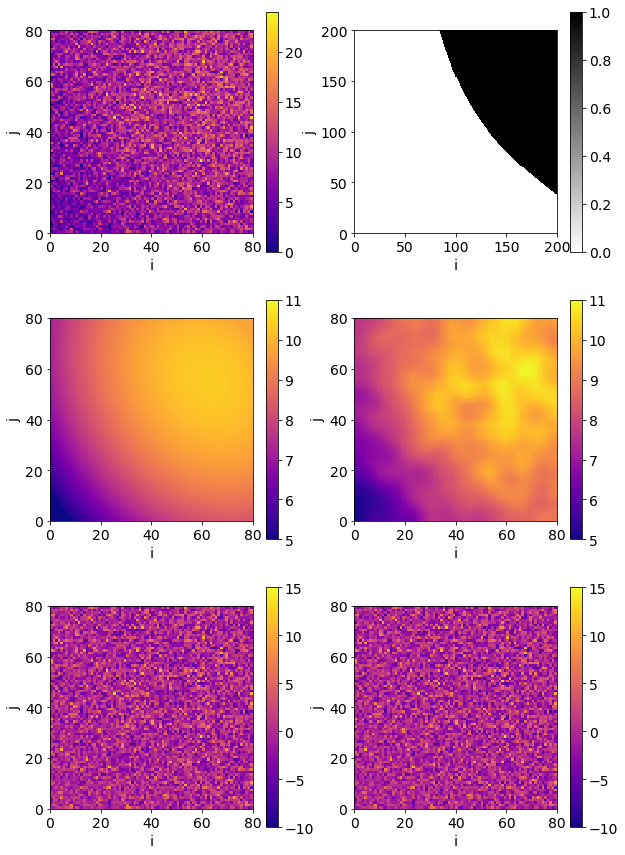}}
\caption{Example of background correction with Earth's presence in the field of view and without bright sources. Top left: Raw shadowgram in counts (20 s exposure). Top right: Earth presence map. Middle left: Fitted model of the shadowgram (2D polynomial fit). Middle right: Wavelet model of the shadowgram ($D_3$). Bottom left: Shadowgram corrected with the fit method. Bottom right: Shadowgram corrected with the wavelet method.}
\label{fig:ex_image_fit_wt}
\end{figure}

Figure \ref{fig:wt_h_sth} gives the distribution of S/N in 20 min sky images for different combinations of the parameters. From this simulation, we propose    using at first  the filter $h_3$ and a maximum scale $s_{\mathrm{th}}=3$ for the cleaning with the wavelet method. The threshold scale $s_{\mathrm{th}}=3$ is also justified by the geometry of the instrument. It is equivalent to keeping, in the image to be deconvolved, the scales whose characteristic size is 1 ($s=0$), 2 ($s=1$), or 4 pixels ($s=2$). The ratio of the mask-element size ($m$) to detector pixel size ($d$) is $m/d= 2.54$. The value $s=2$ is the smallest value for which the characteristic size still covers the mask-element size, while with the value $s=1$ part of the point-source signal would remain in the large scales that are subtracted and would be lost in the image to be deconvolved. The two parameters can be modified after the launch by a telecommand from the ground, based on the observation of the background in flight conditions. 

Figure \ref{fig:ex_image_fit_wt} gives an example of the background correction in the presence of the Earth in the field of view, using either the traditional fit method or the wavelet method. The Earth in the field of view leads to a non-uniform distribution of the counts on the detector, including a lack of counts in the lower left corner (where no part of the detector is totally obscured). Both the traditional fit method and the wavelet method make it possible to correctly model this non-uniformity and to obtain a flat distribution after subtraction.  

\subsection{Galactic ridge X-ray emission non-correction}

\begin{figure}
\resizebox{\hsize}{!}{\includegraphics{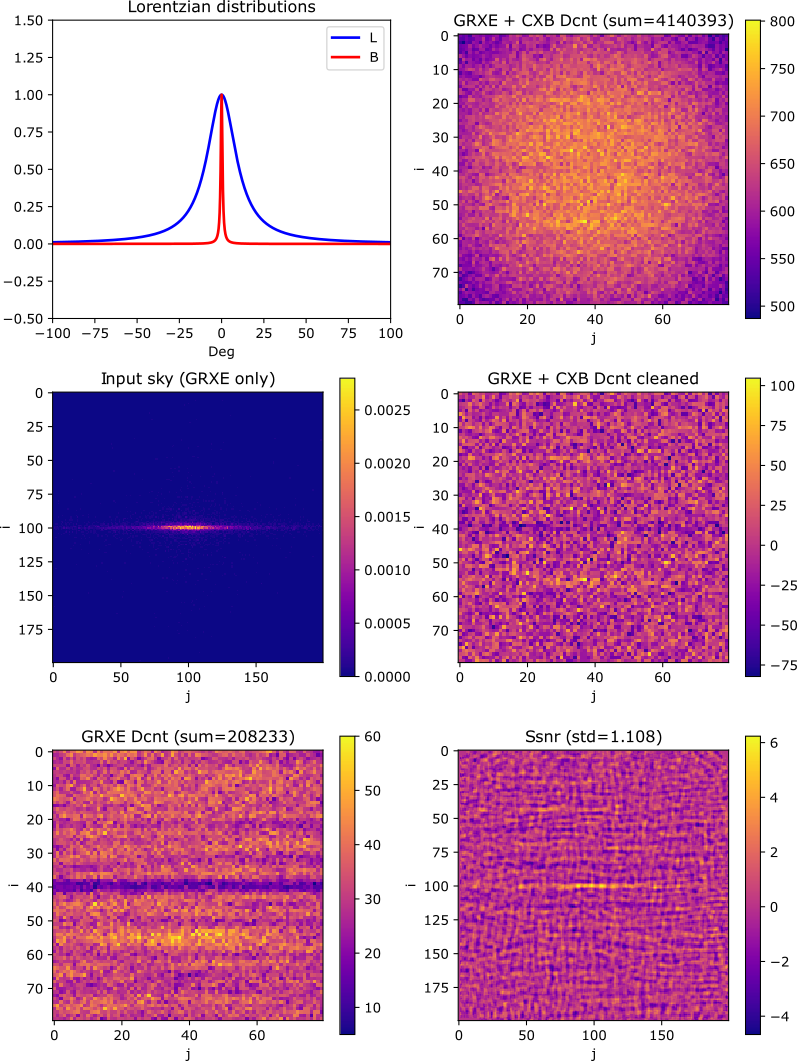}}
\caption{Study of the GRXE in 20~min (scale $n=6$) in 4--10 keV. Top left: The two Lorentzian functions used to model the GRXE in Galactic longitude and latitude. Middle left: Sky map of the points sources used to simulate the GRXE diffuse emission. Bottom left : Shadowgram in counts on the detector (only GRXE). Top right : Shadowgram in counts (GRXE and CXB combined). Middle right: Shadowgram after correction by quadratic fit. Bottom right: Reconstructed sky image in S/N.}
\label{fig:grxe}
\end{figure}

In this section we study the influence of the Galactic ridge X-ray emission (GRXE) and show that its influence on the images produced by ECLAIRs is negligible, and that it is therefore not necessary to integrate it into our simulations nor to worry about its correction.
The GRXE is modelled by a diffuse emission region centred on the Galactic centre. Its spatial distribution is given by two perpendicular Lorentzian functions (Fig. \ref{fig:grxe}, top left) with a full width at half maximum of $21\deg$ along the Galactic plane and $1.2\deg$ perpendicularly \citep{turler_integral_2010}. 
We modelled its spectrum, using as input the distribution shown in \cite{turler_integral_2010} (Fig. 10), which we approximate by a simple power law $E \cdot F_E = K \cdot E^{\alpha}$. Using the two points with coordinates ($E=4$ keV, $E \cdot F_E=10$ keV/cm$^2$/s/sr) and ($E=100$ keV, $E \cdot F_E=1$ keV/cm$^2$/s/sr), we derive the values for the power-law index $\alpha=-1.102$ and the normalization $K=46$ keV/cm$^2$/s/sr at 1 keV. This spectrum is given in a region of $60\times30\deg^2$, which corresponds to 0.54 sr. Hence, the total GRXE flux in that region is $0.549$ ph/cm$^2$/s in 4--10 keV. 
We mimic the diffuse GRXE by 10000 point sources, distributed in the sky according to the two Lorentzian functions described above, where each source has a flux equal to 1/10000 of the total GRXE flux.

In a first simulation (see Fig. \ref{fig:grxe}) the Galactic plane was aligned with one of the cross-arms of the coded mask\footnote{For mechanical reasons, the coded mask has no holes on the two perpendicular central strips forming what is called the `mask cross'. This cross is shown in Figure 1 in \cite{cordier_b_svom_2019} }. As a result, we observe that in 4--10 keV the GRXE projects about 160 counts/s on the detector, while the CXB projects about 3000 counts/s. This case where the Galactic plane is aligned with a mask cross-arm is the worst;   the superposition of the GRXE source projections casts the shadow of one of the cross-arms in the shadowgram (see bottom left panel of Fig. \ref{fig:grxe}). However, in the reconstructed sky images the GRXE is barely visible. Even in the longest exposure of 20 min, the reconstructed structure does not correspond to a point source, the S/R maximum is just above 6$\sigma$, and the standard deviation of the S/N rises to 1.1 (see bottom right panel of Fig. \ref{fig:grxe}), such that this structure   remains below the trigger thresholds even in 20 min exposures.

\begin{figure}
\resizebox{\hsize}{!}{\includegraphics{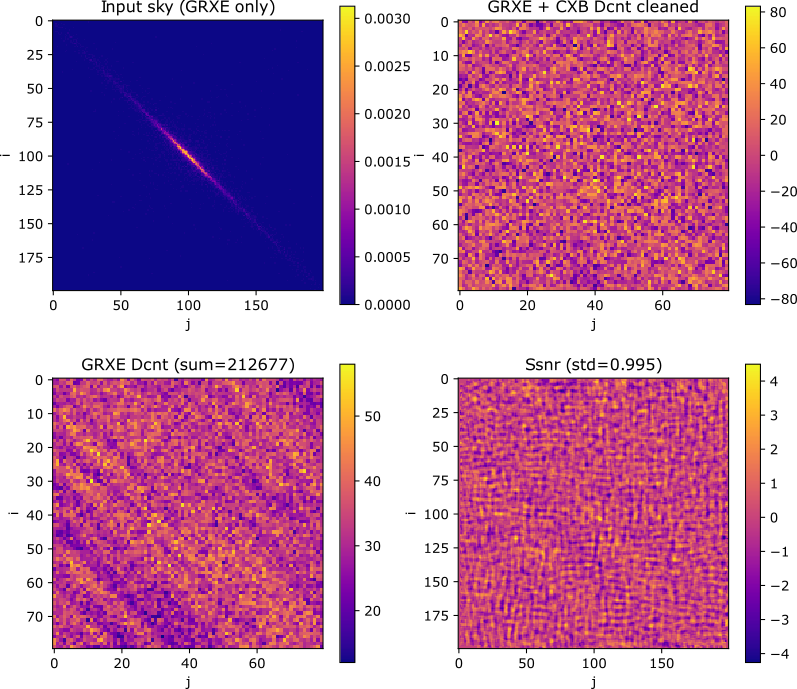}}
\caption{Same as middle and bottom part of Fig. \ref{fig:grxe}, but with the Galactic plane tilted by $45\deg$ inside the field of view.}
\label{fig:grxe_roll}
\end{figure}

We repeated the simulation with the Galactic plane tilted by $45\deg$ in the field of view (Fig. \ref{fig:grxe_roll}). In this case the projections of the simulated sources are no longer aligned with a cross-arm of the mask, which results in a detector shadowgram showing low-amplitude structures only, and a reconstructed sky image where the GRXE diffuse emission is not visible, even for exposures up to 20~min.

Therefore we conclude that the GRXE is negligible in the ECLAIRs image trigger, mainly because the counts produced by the GRXE emission region is smaller than that of the CXB, and because the GRXE is a region of diffuse emission, much larger in size than  can be considered as a point source for ECLAIRs for which the mask imaging is not a completely   appropriate reconstruction technique. Consequently, this emission is not taken into account in our one-year simulation nor in the image trigger software.

Generally speaking, it should be noted that the data from ECLAIRs could be used to study CXB and GRXE via specific ground-based analyses.

\subsection{Correction of known sources}

For each pointing position of the spacecraft, we can build a catalogue of bright X-ray sources and compute their coordinates in the ECLAIRs local frame. The sources that need to be corrected are mainly bright X-ray binaries \citep{dagoneau_onboard_2021}. 

The correction of these known sources in the onboard catalogue is done in the shadowgram by fitting a model of the sources on the shadowgram and subtracting it. The model consists of the weighted sum of the illumination functions of the sources in the field of view (and not obscured by the Earth).
The illumination function of a source is an array of the same size as the detector, which gives, for each pixel of the detector, the pixel fraction illuminated by this source for the given position in the sky.

Depending on how the background is corrected, the fit is either performed simultaneously with the 2D polynomial shape of the CXB (adding one parameter per source in Eq. \ref{eq:2d_shape}) or after the wavelet cleaning (because the projection of the coded-mask pattern on the detector by point sources predominantly contributes  on  small scales).

\begin{figure}
\resizebox{\hsize}{!}{\includegraphics{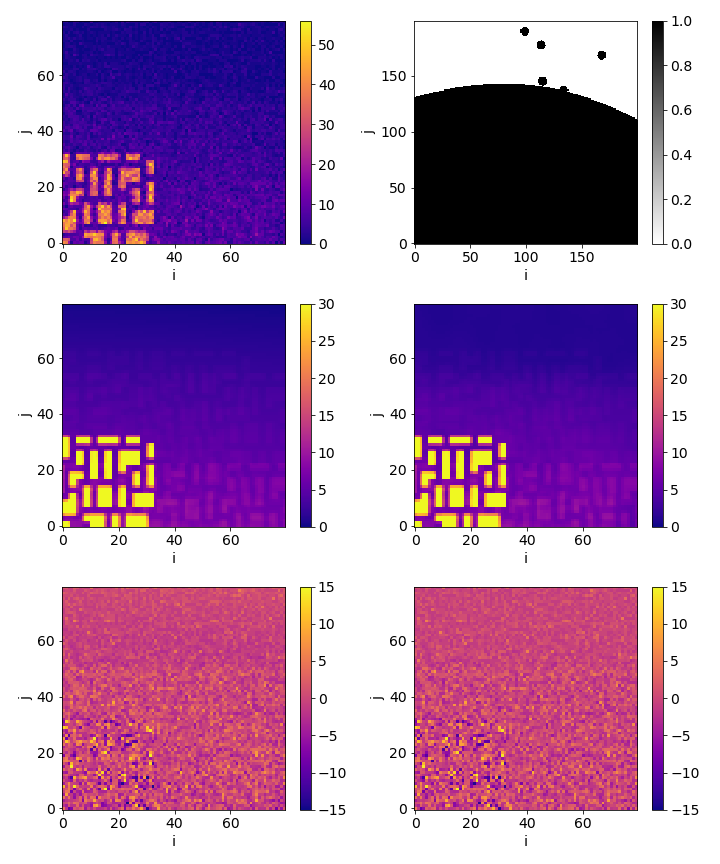}}
\caption{Example of background correction with Earth's presence and with bright sources in the field of view. Top left: Raw shadowgram in counts (20 s exposure). Top right: Sky map with the Earth and the corrected sources  (Sco X-1, GX 349+2, GX 3+1, 4U 1705-440, GX 9+9). Middle left: Fitted model of the shadowgram (2D polynomial and illumination functions of sources). Middle right: Wavelet model of the shadowgram ($D_3$) combined with fitted model of the sources. Bottom left: Shadowgram corrected with the fit method. Bottom right: Shadowgram corrected with the wavelet method.}
\label{fig:ex_image_fit_wt_src}
\end{figure}

Figure \ref{fig:ex_image_fit_wt_src} gives an example of the background correction with Earth's presence and bright sources in the field of view for the two methods. Here the field of view contains many bright sources, including the very bright Scorpius X-1 whose mask shadow is clearly visible in the bottom left corner of the detector image. In the model of the detector, the contribution from the CXB is not visible because the sources, and especially Scorpius X-1, are dominant. After the correction (for both methods), we can see that some counts from the very bright source Scorpius X-1 remain visible in the bottom left corner of the shadowgram. This is caused by the subtraction of an average level from a count distribution that has a large amplitude of variation in the case of a very bright source. However, the consequence is limited because most of the counts are subtracted and a possible residual at the position of the source in the sky will be masked when searching for a new source.

\begin{figure}
\resizebox{\hsize}{!}{\includegraphics{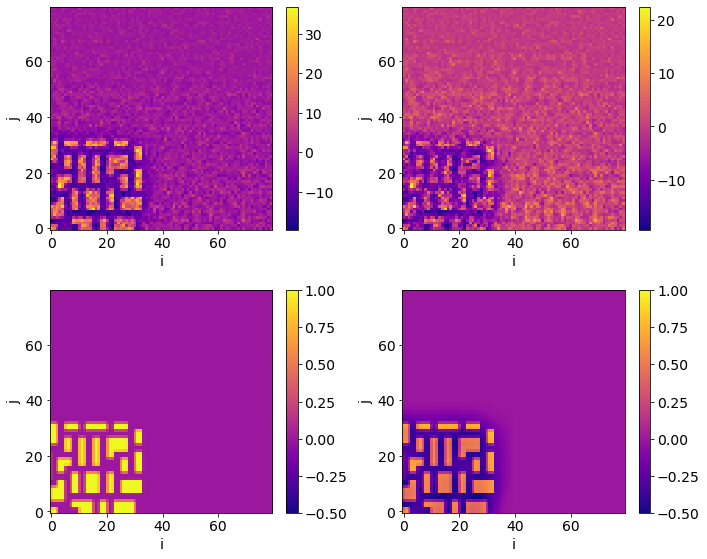}}
\caption{Processing details on the field of view shown in Fig. \ref{fig:ex_image_fit_wt_src}. 
Top left: Shadowgram corrected for the background contribution using the wavelet method without source processing; the contribution of the bright source Sco X-1 remains visible. 
Top right: Shadowgram corrected for the background using the wavelet method followed by the direct fit of the illumination function of the bright known sources (the same non-transformed illumination functions as used for the simultaneous fit with the CXB); the contribution of the bright source is only partially removed by this fit. 
Bottom left: Illumination function (non-transformed)  for a source in the partially coded field of view (used for the direct fit).
Bottom right: Transformed illumination function by the \textit{à trous} wavelet-filter; this is the function to be fitted after the wavelet correction of the CXB in order to  correctly remove the bright source contribution (as shown in  Fig. \ref{fig:ex_image_fit_wt_src}, bottom right).}
\label{fig:ex_image_fit_wt_bad_src_model}
\end{figure}

In the case where the CXB is cleaned with wavelets, the subtraction of the contribution from the known point sources is performed in a second step where the fit is based only on the models of the sources. Since the image has been transformed (Fig. \ref{fig:ex_image_fit_wt_bad_src_model}, top left), the models of the sources to be fitted must also be transformed using the \textit{à trous} algorithm with the same parameters (Fig. \ref{fig:ex_image_fit_wt_bad_src_model}, bottom row) to avoid that the source contributions are only partially subtracted (Fig. \ref{fig:ex_image_fit_wt_bad_src_model}, top right).

\section{After deconvolution: Calibrating the detection threshold}
\label{sec:analyse}

After the cleaning of the detector images from the CXB and from the known bright-source contributions, the shadowgrams of the one-year simulation are ready to be deconvolved in order to reconstruct the sky images.

The sky pixels that are obscured by the Earth are set to 0 in the counts images $S_{\mathrm{cnt}}$, variance images $S_{\mathrm{var}}$, and S/N images $S_{\mathrm{S/N}}$ (only in the first scale of 20.48 s processed by the image trigger).
In order to prevent false triggers from a residual at the position of a corrected catalogued source, pixels within a radius of 6 pixels (configurable, see \citealt{dagoneau_onboard_2021}) around the position of the source are set to 0 in these images. For a source in the centre of the field of view, the pixels within a radius of 6 pixels cover a solid angle of about 0.01 sr. Some other sources, such as Hercules X-1,   which are not bright enough to be corrected in the 20 s shadowgrams but are bright enough to appear in the images of the sky in 20 min, are also masked with zeros at all timescales. 
In order to prevent some artefacts from the deconvolution in very partially coded directions of the sky, the sky pixels for which a mask pattern is projected in less than 50 pixels (configurable) of the detector are also set to 0 in these images. 
This way, pixels with a S/N of exactly 0 are excluded for the computation of the standard deviation and for the new source search by the image trigger.

\begin{figure}
\resizebox{\hsize}{!}{\includegraphics{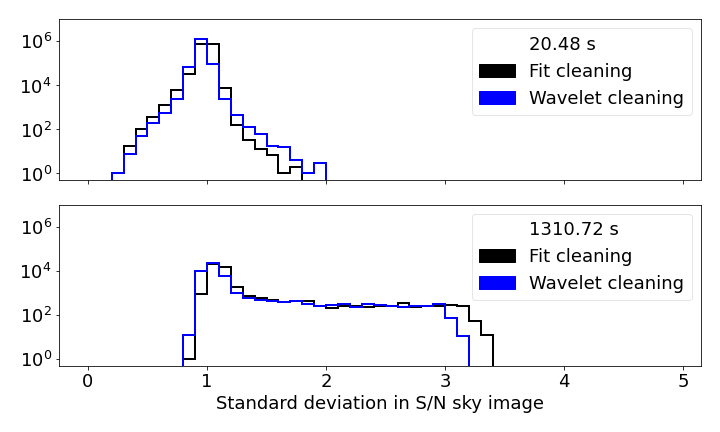}}
\caption{Distribution of the standard deviation for each S/N sky image of 20 s (top, 1510398 entries) and 20 min (bottom, 44895 entries) of the one-year simulation. The black and  blue curves correspond to the images where the CXB was cleaned using the fit method and the wavelet method, respectively.}
\label{fig:histo_std}
\end{figure}

From the previous work we conducted \citep{dagoneau_onboard_2021}, the standard deviation of the S/N in 20 min sky images reaches, without any corrections, values higher than 50 for a pointing towards the Galactic centre and in the region of Scorpius X-1. In addition, with the correction, near the Galactic poles the correction reduces the standard deviation of the S/N in the sky images to values close to 1. Figure \ref{fig:histo_std} gives the distribution of all the standard deviations in S/N images produced during the one-year simulation (in 20 s and 20 min). In 20 s it remains close to 1 in all the produced images, but in 20 min it increases because of the more important number of bright X-ray sources, even if they are corrected, in the Galactic centre. The histograms only includes images composed of ten significant pixels at least, excluding images almost totally obscured by the Earth. The extension to values below and above 1 corresponds to sky images where the Earth occupies a large part of the field of view, and thus few pixels are used to calculate the standard deviation. In the 20 min plot, there are no  values much lower than 1 because in 20 min there are far fewer pixels that are permanently masked by the Earth. 

If the standard deviation allows  the deviation of the S/N to be characterised from the normal distribution, the value that needs to be examined is the maximum value in the  S/N images. The image trigger looks for the maximum value in the S/N images and starts an alert sequence if that value is above a threshold. For a normal S/N distribution the threshold has been set to 6.5 in order to comply with the constraint of one false alert per day as required by the alert system. This value can be obtained from simulations in which a large number of realisations of the background are simulated and corrected;  the sky images are then produced by deconvolution and the maxima in S/N (or rather maxima in S/N divided by the standard deviation in the image in S/N, see below) are recorded. Integrating the distribution of these values from the right until we reach the desired false alarm rate (typically one per day), we find a threshold close to 6.5.  This value is a little higher than the expected theoretical value (4.5) for 400000 independent and identically distributed sky pixels according to a normal distribution $\cal{N}$(0,1) with a threshold at 3$\sigma$ per pixel (in reality the sky pixels are not independent and background residuals may remain after the correction).

\begin{figure}
\resizebox{\hsize}{!}{\includegraphics{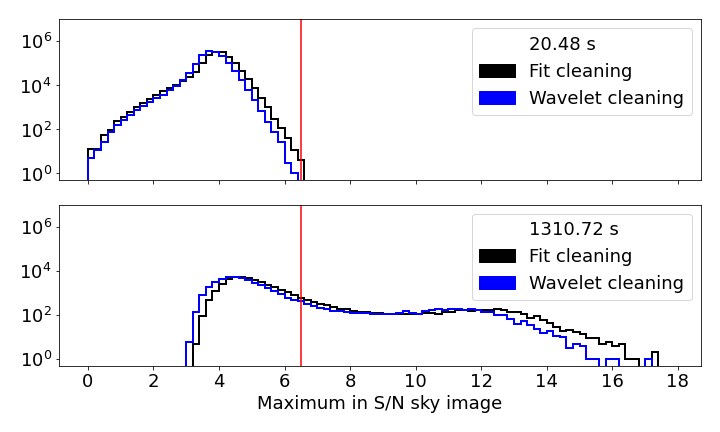}}
\caption{Distribution of the maximum value for each S/N sky image of 20 s (top, 1510398 entries) and 20 min (bottom, 44895 entries) of the one-year simulation. The black and  blue curves correspond to the images where the CXB was cleaned using the fit method and the wavelet method, respectively. The red vertical bar corresponds to a value of 6.5.}
\label{fig:histo_max}
\end{figure}

Thus, we have to limit the presence of values at S/N $>$ 6.5 to keep the threshold of 6.5 and to detect faint transients. Figure \ref{fig:histo_max} gives the distribution of all the maxima in S/N images produced during the one-year simulation (in 20 s and 20 min). In 20 s, the distribution is compatible with a threshold in S/N of 6.5, where there are only a few images with S/N $>$ 6.5. However, in 20 min, there are many images in which the maximum S/N value is much higher than 6.5. This is mainly the case in the Galactic centre where  there are many sources.

\begin{figure}
\resizebox{\hsize}{!}{\includegraphics{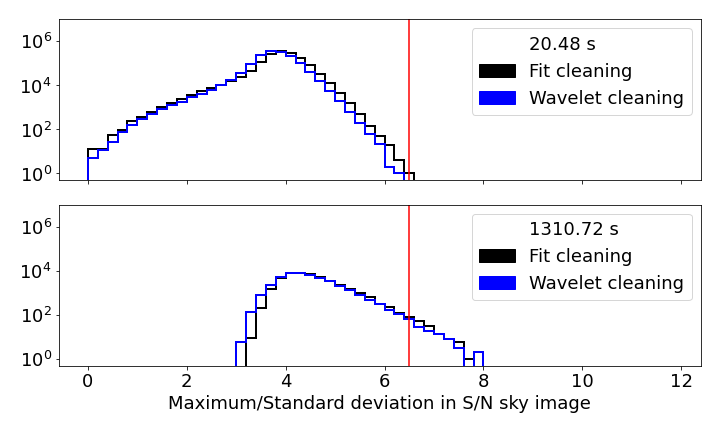}}
\caption{Distribution of the maximum value divided by the standard deviation for each S/N sky image of 20 s (top, 1510398 entries) and 20 min (bottom, 44895 entries) of the one-year simulation. The black and  blue curves correspond to the images where the CXB was cleaned using the fit method and the wavelet method, respectively. The red vertical bar corresponds to a value of 6.5.}
\label{fig:histo_maxOvStd}
\end{figure}

We  also note that the two methods used to correct the CXB give similar results in terms of maximum values and   standard deviation in the S/N sky images. This is the proof that wavelets are able to correct the CXB, without any assumption on the shape of the CXB on the detector.

In order to dynamically adapt the detection threshold according to the quality of the sky images, we propose  applying a threshold of 6.5 times standard deviation   of the S/N in the sky image (but limited to 6.5 if the standard deviation is less than 1). Figure \ref{fig:histo_maxOvStd} gives the distribution of all the maxima divided by the standard deviation in the S/N images produced during the one-year simulation (in 20 s and 20 min). If the standard deviation is smaller than 1, the threshold remains at a value of 6.5. In 20 s, the distribution is not modified very much (because the standard deviation is close to 1). However, in 20 min, the spread of the distribution is reduced. It still overflows beyond 6.5, and thus we can introduce a small dependence of the threshold on the timescale between 6.5 in 20 s and 8 in 20 min. 

\begin{figure}
\resizebox{\hsize}{!}{\includegraphics{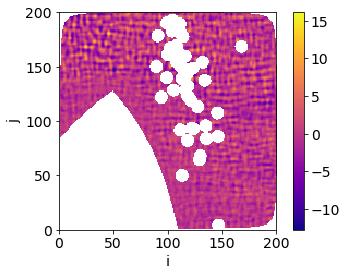}}
\caption{Sky image in S/N in 20 min. White pixels are excluded for the search of new sources: known sources plus part of the field of view that is always obscured by the Earth during the 20 min. The CXB is corrected using the wavelet method.}
\label{fig:ex_snr_image}
\end{figure}

\begin{figure}
\resizebox{\hsize}{!}{\includegraphics{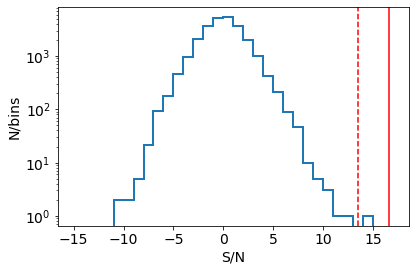}}
\caption{Distribution of the S/N is the sky image of Fig. \ref{fig:ex_snr_image}. The red dashed line and the red plain line correspond to a threshold of 6.5 $\times$   and 8 $\times$ the standard deviation, respectively.}
\label{fig:ex_snr_image_histo}
\end{figure}

\begin{figure}
\resizebox{\hsize}{!}{\includegraphics{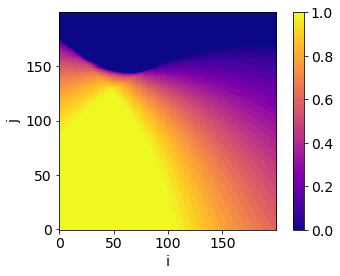}}
\caption{Exposure map for the 20 min corresponding to the sky image shown in Fig. \ref{fig:ex_snr_image}. The value 1 corresponds to the regions that are always obscured by the Earth, the value 0 to those that are never obscured.}
\label{fig:shape_earth}
\end{figure}

We show as an example the longest timescale considered, the 20 min sky image in Fig. \ref{fig:ex_snr_image}, with the same pointing as for Fig. \ref{fig:ex_image_fit_wt_src}. In this image, away from the Earth and the positions of known sources, the S/N standard deviation is 2.08, which sets an adapted detection threshold for new sources of 6.5 $\times$ 2.08 $\approx$ 13.5. However, as we show in the distribution of Fig. \ref{fig:histo_maxOvStd}, this can be too low (the maximum in the sky image is 14.1) and raises false alerts. Increasing  the threshold slightly  to 8 $\times$ 2.08 $\approx$ 16.6 allows detection at values that are far enough from the bulk of the distribution (see Fig. \ref{fig:ex_snr_image_histo}).

In Fig. \ref{fig:ex_snr_image} the shape of the Earth may be surprising. This is caused by the fact that the part of the sky masked by the Earth is replaced by zero values only in the 20 s images. Thus, in the sky images summed up to 20 min, only the part of the sky continuously occulted by the Earth is masked. The regions that the Earth has only partially occulted during the 20 min remain unmasked. Figure \ref{fig:shape_earth} gives the exposure map for these 20 min, and shows the portion of the sky permanently obscured by the Earth and those that are only obscured during a fraction of the exposure time. 

\begin{figure}
\resizebox{\hsize}{!}{\includegraphics{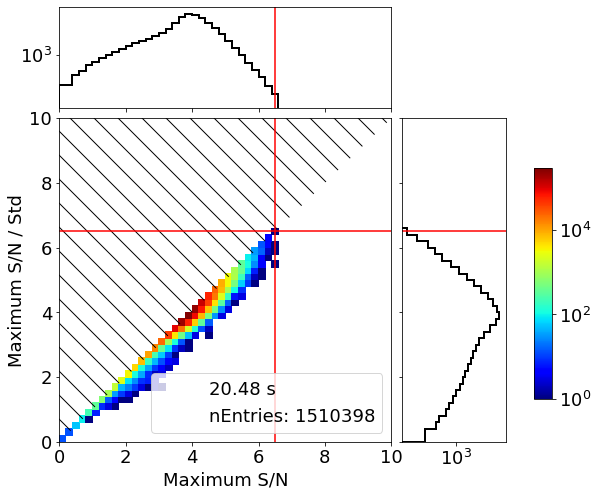}}
\caption{Two-dimensional distribution of the maximum S/N vs the maximum S/N divided by the standard deviation for each sky image of 20 s. The colour corresponds to the number of entries (sky image) in each bin. The red lines show the threshold applied on each axis. The background is corrected with the fit method. The hatched area corresponds to a region that is never populated because the standard deviation used for division is never smaller than 1.}
\label{fig:plot_2D_scale_0_fit}
\end{figure}

\begin{figure}
\resizebox{\hsize}{!}{\includegraphics{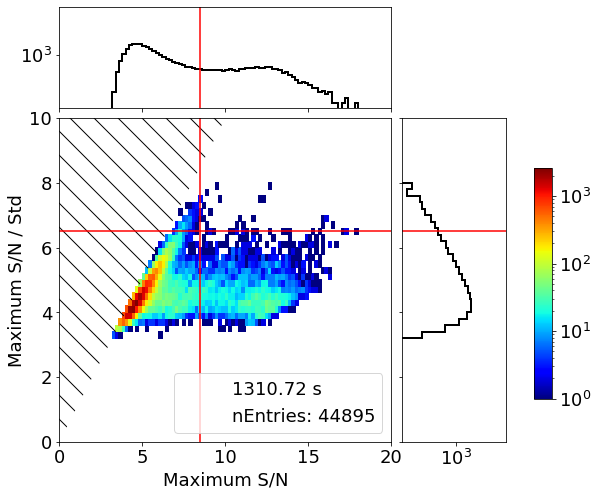}}
\caption{Same as Fig. \ref{fig:plot_2D_scale_0_fit}, but for the timescale of 20 min.}
\label{fig:plot_2D_scale_6_fit}
\end{figure}

Finally, instead of considering separately the distributions of the measured quantities in the S/N images, the maximum and the maximum over the standard deviation, we now determine the 2D distribution of the two quantities, shown in Fig. \ref{fig:plot_2D_scale_0_fit} for the timescale of 20 s and in Fig. \ref{fig:plot_2D_scale_6_fit} for the timescale of 20 min. 
On the short timescale, both distributions are largely correlated, which permits  setting the thresholds in order to limit the false alert rate to the value of 6.5 for both measured quantities.
On the large timescale, the 2D distribution is much larger for the maximum, while the spread in the maximum over standard deviation is mainly observed for low values of the maximum. 
This permits   lowering the thresholds for both quantities, and requiring that the maximum must be above the value of 8.5 and at the same time the maximum over standard deviation   above the value of 6.5. This logical `and' condition considers  the events located in the upper right corner of the figure to be acceptable, and leaves  more room for detection than with the more stringent thresholds derived above with the distributions of the two quantities considered separately.

\section{Conclusion}

In this paper we have presented the algorithm behind the image trigger on board \textit{SVOM}/ECLAIRs. This trigger will build and analyse images of the sky on timescales of 20 s and 20 min. We described the two methods that are implemented to correct the CXB background: a traditional fit method that has to work in the case of the Earth's presence in the field of view, and an alternative method based on the \textit{à trous} wavelet decomposition. The two methods give similar results in terms of correction of the detector images that are built from a one-year simulation of the background, including known X-ray sources from the catalogue we developed previously. In addition, we performed a calibration of the detection threshold that adapts to the quality of the reconstructed sky image according to the standard deviation of the S/N in the sky image, with a weak dependence on the timescale.

The traditional method of correcting the background by fitting a 2D shape will be used from the beginning of flight operations, while the wavelet-based method, never applied before the deconvolution for coded-mask aperture telescopes, will be validated on the ground with the first data. The background correction with the wavelet method does not require any assumption on the shape of the background. Thus, it could be more resilient, for example towards the spatial variations of the background on the detector when the Earth   crosses the field of view or following the activation of the detector by high-energy particles during passages through the South Atlantic Anomaly of the Earth's magnetic field.

The different steps that we described will be implemented in the onboard software. 
The code is executed on the onboard Leon3 processor, and part of it has been benchmarked with a few reference cases for computing performance and compared to execution times obtained on a Linux machine from which we derive the following estimates.
Without sources in the field of view, the background correction takes about 0.4 s for the polynomial fit of the CXB and the wavelet background correction takes about 0.3 s. With ten sources present, the combined fit of the CXB and the sources takes about 1.1 s, while the wavelet background correction still takes 0.3 s to which 0.6 s need to be added for the fit of the ten sources alone. To these values, about 0.9 s need to be added for the sky deconvolution. It appears that the wavelet method is therefore slightly faster than the background correction by combined fit in all cases.

Once ECLAIRs is in flight, the same code   implemented on board can also be executed on ground computers using all the photons recorded by the ECLAIRs camera, which will be sent to the ground through X-band with 6 h to 12 h of delay. Thus, the algorithm of the trigger and the different methods for correcting the background can be executed offline on real data to test the various parameters and to fine-tune them, in order to optimise the sensitivity of the instruments towards faint X-ray transients.    

\begin{acknowledgements}
ECLAIRs is a cooperation between CNES, CEA and CNRS, with CNES acting as prime contractor. This work is supported by CEA and by the “IDI 2017” project of the French “Investissements d’Avenir” program, financed by IDEX Paris-Saclay, ANR-11-IDEX-0003-02. The authors would like to warmly thank the anonymous referee for very helpful comments and suggestions, including the question about the influence of the Galactic Ridge X-ray Emission on long duration image reconstruction.
\end{acknowledgements}

\bibliographystyle{aasjournal}
\bibliography{references}  

\end{document}